\documentclass[aps,twocolumn,nopacs,floats,superscriptaddress]{revtex4-1}
\usepackage{graphicx}
\usepackage{amssymb,amsmath}
\usepackage{color}
\usepackage{bm}
\newcommand{\beq}{\begin{equation}}
\newcommand{\eeq}{\end{equation}}
\newcommand{\bea}{\begin{eqnarray}}
\newcommand{\eea}{\end{eqnarray}}

\begin{document}

\author{Andrey Chubukov}
\affiliation{Department of Physics, University of Minnesota, Minneapolis, MN 55455, USA}
\author{Nikolay V. Prokof'ev}
\affiliation{Department of Physics, University of Massachusetts, Amherst, MA 01003, USA}
\affiliation{National Research Center ``Kurchatov Institute," 123182 Moscow, Russia}
\author{Boris V. Svistunov}
\affiliation{Department of Physics, University of Massachusetts, Amherst, MA 01003, USA}
\affiliation{National Research Center ``Kurchatov Institute," 123182 Moscow, Russia}
\affiliation{Wilczek Quantum Center, School of Physics and Astronomy, Shanghai Jiao Tong University, Shanghai 200240, China}

\title{Implicit renormalization approach to the problem of Cooper instability}

\date{\today}
\begin{abstract}
In the vast majority of cases, superconducting transition takes place at exponentially low temperature $T_c$
out of the Fermi liquid regime. We discuss the problem of determining $T_c$ from known system properties
at temperatures $T \gg T_c$, and stress that this cannot be done reliably by following the standard
protocol of solving for the largest eigenvalue of the original gap-function equation. However,
within the implicit renormalization approach, the gap-function equation can be used to formulate an alternative
eigenvalue problem, solving which
 leads to an accurate prediction for both $T_c$
and the gap function immediately below $T_c$.
With the diagrammatic Monte Carlo techniques,
this eigenvalue problem can be  solved without invoking the matrix inversion or
even explicitly calculating the four-point vertex function.
\end{abstract}


\maketitle

\section{Introduction}
\label{sec:Introduction}


A conventional Bardeen-Cooper-Schrieffer (BCS) theory of $s$-wave superconductivity and its extensions to other pairing symmetries, and to strong coupling, assume that the pairing interaction is attractive in at least one pairing channel.  In the BCS theory, the attraction comes from phonon exchange, in  theories of non-ordinary $s$-wave superconductivity (e.g., $d$-wave superconductivity in the cuprates, $s^{+-}$ superconductivity in Fe-based metals, etc.), the attractive interaction between fermions is believed to originate from screened Coulomb interaction between electrons. In itinerant models of the electronic pairing, the attraction comes from an exchange of collective bosonic excitations in either  spin channel (the  spin-fluctuation exchange for superconductivity near the onset of magnetism~\cite{Pines,Scalapino2012,Abanov2003,Mazin2008a,
        Kuroki,Chubukov2008}) or in charge channel (e.g., the exchange
 of nematic fluctuations near the onset of a nematic order~\cite{Fradkin2010,Lederer2015,Lederer2017,Berg2019,Klein2018}).

Recent interest in $s$-wave superconductivity in  SrTiO$_3$~~\cite{Schooley64,Schooley65,Lin14,Triscone2017},  Pb$_{1-x}$Tl$_x$Te~~\cite{Chernik1981}, half-Heusler compounds~\cite{Nakajima2015}, and single-crystal Bi~\cite{Prakashaaf8227} re-ignited the discussion of another aspect of the pairing problem: the interplay between a weaker electron-phonon interaction and a stronger electron-electron repulsion~\cite{Gurevich1962,Schooley64,Schooley65,Chernik1981,Takada1980,Ikeda1992,Grimaldi1995,Mahan,Lin14,
Nakajima2015,Prakashaaf8227,Edge2015,Ruhman2016,Gorkov2016,Gorkov2017,Ruhman2016,Ruhman2017, DHLee2015,Rademaker2016,Zhou2016,Zhou2017,
   Trevisan2018,Savary2017,Rowley2018,Lonzarich2018,Woelfle2018,Sadovskii2018,Sadovskii2018b,Aperis2018,Aperis2018b}.
BCS theory (and its finite-coupling version, known as Eliashberg theory~\cite{Eliashberg,Migdal,Scalapino1969,Carbotte90,Carbotte_Marsiglio}) neglects electron-electron repulsion and considers only the phonon-mediated attraction.
It has been argued long ago~\cite{Tolmachov2,Morel1962,Scalapino66,McMillan1968,Scalapino1969,Carbotte90} that in a situation, when the Debye frequency $\Omega_D$ is much smaller than
the Fermi energy $E_F$ (or, more precisely, the characteristic energy $E_c \leq E_F$,
up to which one can expand the dispersion linearly near the Fermi level),
  the repulsive Coulomb interaction gets logarithmically suppressed by fermions with energies between $E_c$ and $\Omega_D$, and electron-phonon attraction, emerging at energies below $\Omega_D$, wins over the reduced Coulomb repulsion.  The actual situation is more tricky, however,  because the net interaction---the sum of the Coulomb repulsion and the electron-phonon interaction---is repulsive, and remains repulsive at all energies.  The key effect of the electron-phonon interaction is that it makes the effective pairing interaction $V_{\rm eff} (\omega)$ dependent on the transferred frequency, interpolating
 between a smaller value at $\omega \to 0$  and a larger value at $\omega > \Omega_D$~\cite{Gurevich1962,Scalapino66,McMillan1968,Takada1980,Coleman_book,Ruhman2016,Ruhman2017}.
 A simplified model of such an interaction has been considered by
Rietschel and Sham~\cite{Rietschel1983,Coleman_book}.  They replaced the actual frequency-dependent
$V_{\rm eff} (\omega = \omega_m - \omega_n)$  along the Matsubara axis by a step-like
 $V_{\rm RS}
 (\omega_m, \omega_n)$ with separable dependence on $\omega_m$ and $\omega_n$:
\begin{equation}
V_{\rm RS} = \left\{ \begin{array}{l}
0, \quad \mbox{if} \;\: |\omega_n|> E_c, \; \mbox{or} \; |\omega_m| > E_c ,\\
\\
g - gf \theta(\Omega\!  -\! |\omega_n|) \theta(\Omega \! -\! |\omega_m|)  \;\; \mbox{otherwise} ,
\end{array}
\right.
\label{ModelA1}
\end{equation}
where $g$ is positive and  $0<f<1$ [i.e., the interaction (\ref{ModelA1}) is of repulsive character].  This $V_{\rm RS}$ has three values:
 $g (1-f)$ at small frequencies, $g$ at larger frequencies, and zero at very high frequencies. In the limit when $l = \ln [E_c/\Omega]$  is large, the analysis of the linearized gap equation shows~\cite{Rietschel1983,Coleman_book} that $T_c$ is finite when $f > 1/(1+ g l)$ (see below).  At large enough  $l$  this holds for any $f >0$, i.e., for any interaction, which  gets reduced below $\Omega$.  The contribution of the ``average" repulsion $g (1-f/2)$ to the gap equation is eliminated by sign change of the gap function $\Delta (\omega_n)$ between $\omega_n <\Omega$ and $\Omega < \omega_n < E_c$, much like
   the on-site Hubbard repulsion
    gets eliminated from the gap equation
    for $s^{+-}$ superconductivity~\cite{Chubukov_Maiti}.

This paper has two goals.  First, we want to analyze superconductivity for more realistic  repulsive interaction.  Several recent studies of superconductivity in SrTiO$_3$ and Bi argued that $V_{\rm eff} (q,\omega)$ can be viewed as the  sum of a screened Coulomb interaction and an interaction with a gapped boson, dressed by the Coulomb potential, where a boson is a hybridized mode between a longitudinal phonon and a plasmon~\cite{Mahan,Ruhman2016,Ruhman2017,Maria2019}. We focus on the frequency dependence of $V_{\rm eff} (q,\omega)$ and neglect its momentum dependence.
Specifically, we consider
\beq
V_{\rm eff}(\omega) = g \left(1 - \frac{\Omega^2_a}{\omega^2 + \Omega^2}\right)  = g \left(\frac{\omega^2 + \Omega^2_1}{\omega^2 + \Omega^2}\right) \,,
\label{ch_1}
\eeq
where $\omega$ is a running transferred Matsubara frequency, $\Omega$ is the frequency of a bosonic mode, and
$\Omega_a \leq \Omega$, i.e.,  $\Omega^2_1  = \Omega^2 - \Omega^2_a < \Omega$.  In the limiting case
$\Omega_a = \Omega$, i.e., $\Omega_1=0$, the model reduces to the modified Bardeen-Pines model~\cite{BardeenPines} (with
 a gapped boson at  frequency $\Omega$ instead of an acoustic mode, as in the original Bardeen-Pines model).

 The pairing interaction $V_{\rm eff} (\omega)$  is similar to $V_{\rm RS} (\omega)$ in the sense that it  reduces to
  a larger
  repulsion $g$ at large frequencies and to a smaller repulsion $g(\Omega_1/\Omega)^2 < g$ at small frequencies. However, in distinction to $V_{\rm RS} (\omega)$, the
$V_{\rm eff} (\omega)$ form smoothly interpolates between the two limits, and
is a function of a single variable $\omega = \omega_m - \omega_n$, rather than a separable function of $\omega_m$ and $\omega_n$.

We analyze superconductivity in this model analytically,  in the weak coupling limit of small $g$. We show that
 the results are qualitatively similar to those of the Rietschel-Sham model
  in that
  at $\Omega_1 =0$, $T_c$ is non-zero for arbitrary weak $g$, and for $0 < \Omega_1 < \Omega$, there is a threshold on $g$, below which superconductivity does not develop. However, the threshold value and the value of $T_c$ above the threshold are different from those in the Rietschel-Sham model.   The absence of a threshold on $g$ at $\Omega_1 =0$, i.e., at $\Omega_a = \Omega$ can be understood as the consequence of the fact that this is a boundary between a repulsive and an attractive interaction: at infinitesimally larger $\Omega_a$ the interaction becomes attractive at the smallest frequencies, in which case $T_c$ is finite at any $g$, like in BCS theory.

In most
of realistic
 cases,
 an analytical solution is not possible  and one has to rely on
a numerical procedure of determining $T_c$
 and the
 gap function immediately below $T_c$.
 This brings us to the second goal of our work---to set up the computational protocol to obtain
critical parameters  in the generic case of frequency and momentum dependent interaction.
In general, one has to solve for the largest eigenvalue $\lambda_{\rm max}$ of the gap-function equation and obtain $T_c$ from the condition $\lambda_{\rm max} =1$~~\cite{Scalapino1969,Gladstone1969}.  In practice, $T_c$ at weak coupling is small, and to reach it, one needs to consider a very large number of Matsubara
points
(and momenta points near the Fermi surface).
 This creates a serious
computational  challenge.
The conventional recipe in this situation would be to
 assume that at a temperature, at which a numeric simulation is done,
  the gap function
  is already saturated to its value in the superconducting state immediately below $T_c$. 
Below we will call this a critical gap function.  Under this assumption,
the functional form, and, correspondingly,
the flow of $\lambda_{\rm max}$ with $T$ is logarithmical, allowing
one to approximate $\lambda_{\rm max}$ as
$\lambda_{\rm max} = g \rho_F \ln [\Omega/T]+{\rm const}$,
where $\rho_F$ is the density of states per spin component. Using this relation, one can  extrapolate $\lambda_{\rm max} (T)$ from $T \gg T_c$, where $\lambda_{\rm max}$ can be obtained with a manageable number of frequency and momentum points,
to a much smaller $T$, and obtain $T_c$ from the condition $\lambda_{\rm max} (T_c) =1$.

We argue that, while this approach works well for the case of an attractive potential, when the gap function does not change sign as a function of frequency, it fails for the case of a frequency dependent repulsive interaction, like
$V_{\rm RS} (\omega_m,\omega_n)$ or $V_{\rm eff} (\omega)$.
The key reason is that for a repulsive interaction, the ratio of the gap function in the frequency range where it is positive, and the one where it is negative, by itself depends on the logarithm of temperature, and this additional logarithmical dependence
cannot be neglected in the extrapolation procedure.
This leads to a non-linear dependence of $\lambda_{\rm max}$ on $\ln [\Omega/T]$,
and makes the extrapolation from $T \gg T_c$ unreliable for determining $T_c$.
We show this explicitly for both the Rietschel-Sham and $V_{\rm eff}(\omega)$ models.

We introduce the new protocol, which overcomes this complication.
We call it an implicit renormalization approach.
Specifically, we re-formulate the eigenvalue problem in such a way that the new largest eigenvalue,
${\bar \lambda}_{\rm max} (T)$, which still obeys ${\bar \lambda}_{\rm max} (T_c) =1$,
remains linear in $\ln [\Omega/T]$
in the whole $T$ range between $\Omega$ and much smaller $T_c$.
We show that this allows one to determine $T_c$ with high precision by
the extrapolation from higher $T$.
Furthermore, the eigenvector of the implicit renormalization protocol
is straightforwardly related to the critical gap function.
 After re-weighting its low-frequency
part with the factor ${\bar \lambda}_{\rm max} (T)$, the former becomes equal to the latter.
 We apply the implicit renormalization method to both the  Rietschel-Sham model and the model
with $V_{\rm eff} (\omega)$ and in both cases find that it works with a remarkable accuracy.

The structure of the paper is the following.  In the next section we present the analytical solutions for $T_c$, first for the exactly solvable Rietschel-Sham model and then for the model with more
 realistic $V_{\rm eff} (\omega)$.  In Sec.~\ref{sec:numerics} we
discuss the extrapolation problem and introduce the implicit renormalization approach to determine $T_c$ in both models.
In Sec.~\ref{sec:newpseudo}, we explain the method in more general terms
and how to apply it to metals, when the
irreducible interaction in the Cooper channel is given by a  four-point vertex $\Gamma$,
which depends on both frequency and momentum deviations from the Fermi surface.
We argue that even a static repulsive interaction, which is weaker near the Fermi surface than away from it,
may give rise to a finite $T_c$.
In this Section we also discuss the Diagrammatic Monte Carlo (DiagMC) method
and explain that to successfully apply the implicit renormalization scheme
one does not need to know $\Gamma$ explicitly---all computed objects are no more
complex than the single-particle Green's function while all integrals and summations
over the diagrammatic space are performed stochastically.
In Sec.~\ref{sec:conclusions}, we provide further discussion and present conclusions.

\section{Analytical solution for  $T_c$ in models with frequency-dependent repulsive interaction}
\label{sec:screening}
\begin{figure}[htb]
\includegraphics[width=0.80\columnwidth]{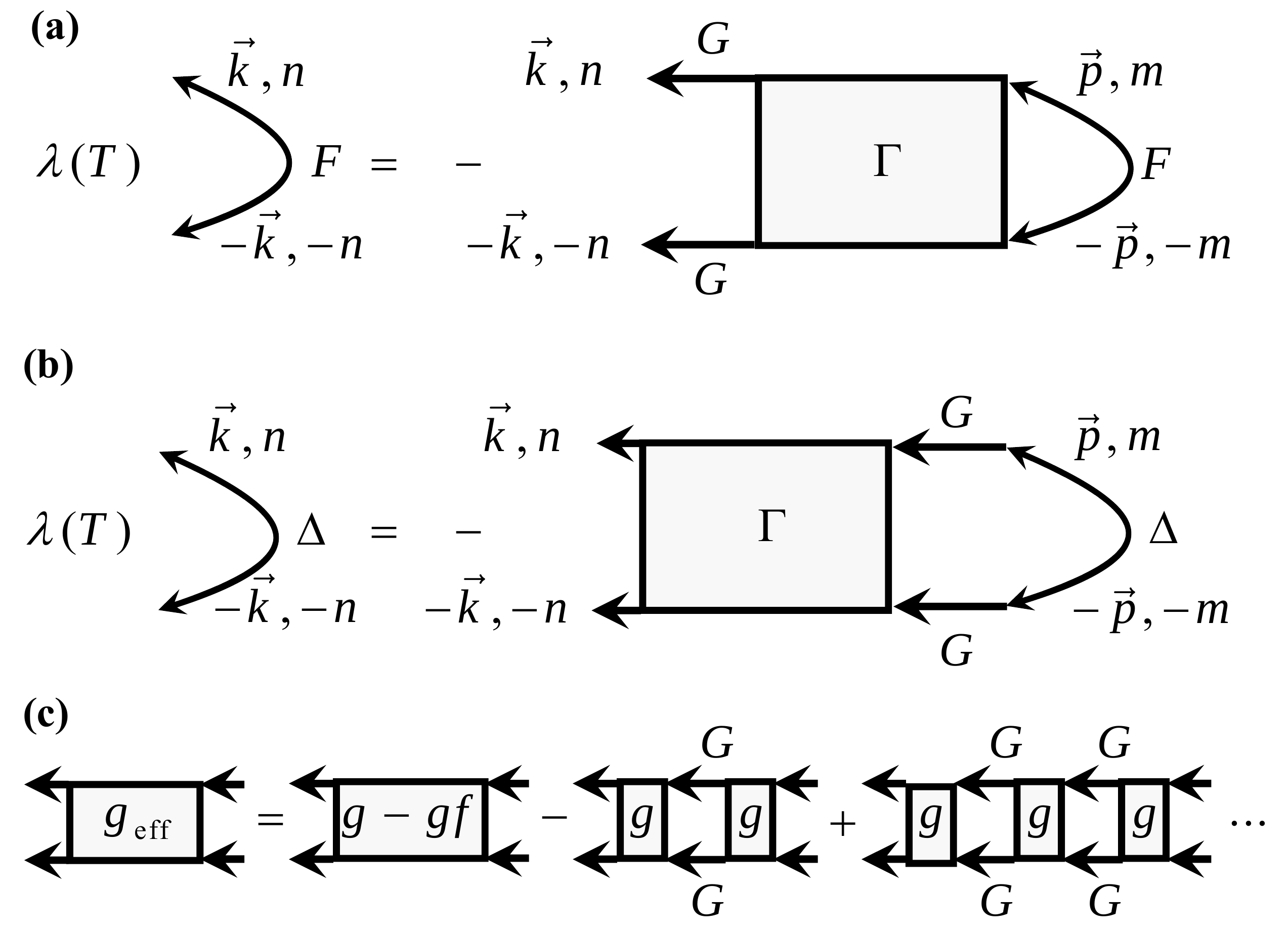}
\caption{ Panels (a) and (b): Graphic representation of the equation for the gap function.
The two eigenvalue equations are identical to each other up to the substitution $F=GG\Delta$.
Panel (c): Equation for the effective low-frequency vertex function $\Gamma_{\rm eff}$ for model (\ref{ModelA1}).
All incoming and outgoing frequencies satisfy $(\omega_n, \omega_m) <\Omega$, while
all intermediate lines are restricted to high frequencies $\omega_k > \Omega$.
\label{fig1}}
\end{figure}

The eigenvalue problem for the gap function $\Delta_{{\mathbf k},n}$ in $d$ dimensions reads
[see Fig.~\ref{fig1}(a)-(b); we use units such that $\hbar=1$ and $k_B=1$]:
\begin{equation}
\lambda(T) \Delta_{{\mathbf k}, \omega_n} = - T \sum_m \int \frac{d{\mathbf p}}{(2\pi)^d} \,
\Gamma^{{\mathbf k},\omega_n}_{{\mathbf p}, \omega_m} \, G^{(2)}_{{\mathbf p}, \omega_m} \; \Delta_{{\mathbf p},\omega_m} \,.
\label{GE1}
\end{equation}
Here $\Gamma $ is a four-point vertex with zero
incoming momentum and frequency, irreducible in the particle-particle (Cooper) channel,  $\omega_n =\pi T(2n+1)$ and $\omega_m = \pi T(2m+1)$ are  fermionic Matsubara frequencies,
 $G^{(2)}$ is the product of two single particle Green's functions,
$G^{(2)}_{{\mathbf p},m} \equiv G_{{\mathbf p},m} G_{-{\mathbf p},-m}$.
To simplify the discussion,
we consider $s$-wave pairing and assume that $\Gamma^{{\mathbf k},n}_{{\mathbf p}, m}$ depends on frequencies
$\omega_n$ and $\omega_m$, but does not depend on momenta; for the Green's function we take the form
$G_{{\mathbf p},m} = 1/(i \omega_m - \xi_p)$ and assume that the quasiparticle dispersion $\xi_p$
can be linearized around the Fermi surface for all $|\xi_p| \leq E_c$.
Under these assumptions,  the momentum integration can be carried out exactly, and the eigenvalue problem reduces to
 \begin{equation}
\lambda(T) \Delta_{\omega_n} = -\pi T \sum_m V(\omega_m, \omega_n) \frac{\Delta_{\omega_m}}{|\omega_m|}
\, .
\label{GE1_1}
\end{equation}

We first consider the exactly solvable Rietschel-Sham model with $V(\omega_m, \omega_n) = V_{\rm RS} (\omega_m, \omega_n)$  and then discuss analytical approach to the model with frequency-dependent repulsive interaction $V_{\rm eff} (\omega_n-\omega_m)$.

\subsection{Rietschel-Sham model}
\label{sec:RS}

The pairing interaction in the Rietschel-Sham model, $V_{\rm RS} (\omega_m,\omega_n)$, is given by Eq.~(\ref{ModelA1}). We recall that this interaction has a step-like form and is separable between $\omega_m$ and $\omega_n$.
The step-like $V_{\rm RS}$ equals
 $g (1-f)$ at small frequencies, $|\omega_m|, |\omega_n| < \Omega$  and equals $g$ at large frequencies, $\Omega < |\omega_m| < E_c, ~\Omega < |\omega_n| < E_c$.

 For such pairing potential, the gap function $\Delta (\omega_n)$ also  displays a step-like behavior and can be parameterized as
\begin{equation}
\Delta (\omega_n) = a\theta(E_c -|\omega_n|) + (b-a)\theta(\Omega -|\omega_n|) \,.
\label{ModelA2}
\end{equation}
The eigenvalue problem is then reduced to solving the quadratic equation for the ratio $a/b$.
 It is convenient to define
\begin{equation}
\ell = \pi T \sum_{E_c >|\omega_m|>\Omega} \frac{1}{|\omega_m|}, ~~L = \pi T \sum_{|\omega_m|<\Omega} \frac{1}{|\omega_m|}
\,.
\label{ModelA3}
\end{equation}
To logarithmic accuracy,
\begin{equation}
L =  \ln [\Omega /\alpha T]  \,, \qquad \ell = \ln [E_c/\Omega] \,.
\label{ModelA4}
\end{equation}
Here $L$ is a conventional Cooper logarithm, for which $\alpha=0.882$ is
the factor that one needs to add to convert the discrete sum $ 2 \pi T \sum_{\pi T}^{\Omega} \omega_m^{-1}$ into the $\int_{\alpha T}^{\Omega} d\omega / \omega $, and $\ell$ (sometimes called Tyablikov-McMillan logarithm) accounts for the
reduction of the repulsive interaction at energies between $E_c$ and $\Omega$.

Using these notations and assuming that both $L$ {\it and} $\ell$ are large, one can cast the eigenvalue problem into a compact form
\begin{eqnarray}
\left\{
{\begin{array}{*{20}{l}}
\lambda a= -g\ell a - gLb ,\\
\lambda b= -g(1-f) L b - g l a .
\end{array}} \right.
\label{ModelA05}
\end{eqnarray}
The transition temperature is determined by the condition $\lambda_{\rm max} (T_c) =1$, where $\lambda_{\rm max}$ is the largest eigenvalue of (\ref{ModelA05}).  Solving for $\lambda_{\rm max}$ we
obtain
\begin{equation}
\lambda_{\rm max} (T) = \frac{g}{2} \left[ \sqrt{(L(1-f)+\ell )^2+4fL\ell } \, -\,  L(1\! -\! f) - \ell \right] \,.
\label{solA}
\end{equation}
We see that $\lambda_{\rm max}$  is {\it positive} and monotonically
increases with decreasing  temperature, as long as $f$ is non-zero, i.e., as long as the interaction varies between small and large frequencies. However, we also see that $\lambda_{\rm max} (T)$ is not simply proportional to $L$, i.e., the increase of $\lambda_{\rm max} (T)$ with decreasing $T$  is not simply logarithmical. Moreover,  at $T=0$, when  $L \to  \infty$, $\lambda_{\rm max}$  saturates at $\lambda_{\rm max} (T=0) = g\ell f/(1-f)$.
When this limiting value drops below unity, i.e., when
\begin{equation}
f <1/(1 + g \ell) \, ,
\label{Tolm}
\end{equation}
the system remains in the normal state down to $T=0$.  This holds when $f$ is non-zero and $l$ is finite. If we fix $f$ and keep increasing $l$, we find that for large enough $l$, i.e., for large enough separation between $E_c$ and $\Omega$,
$T_c$ is still finite. Setting $\lambda_{\rm max} (T_c) =1$ in (\ref{solA}), we obtain
\beq
L = \frac{1+ g \ell}{g} \frac{1}{g \ell f - (1-f)} \,.
\label{ch_7}
\eeq
 The existence of a finite $T_c$ at large enough $\ell$ agrees with the general reasoning that at $E_c/\Omega \gg 1$,   a repulsive Coulomb interaction gets strongly reduced at energies comparable to $\Omega$ and does not overshadows an attraction due to phonon exchange.

\subsection{The model with $V_{\rm eff} (\omega)$}

We next discuss how the results get modified if we replace separable step-like $V_{\rm RS} (\omega_n, \omega_m)$ by
$V_{\rm eff} (\omega)$ from Eq.~(\ref{ch_1}), which is a continuous function, and
depends on the frequency transfer $\omega = \omega_m - \omega_n)$ rather that
  separately on $\omega_m, \omega_n$.

The equation for the gap function
 now
 reads:
\begin{equation}
\lambda  \Delta_n = - g \pi T \sum_m^{|\omega_m|<E_c} \;
\frac{(\omega_n-\omega_m)^2 + \Omega^2_1}{(\omega_n-\omega_m)^2 + \Omega^2}\, \frac{\Delta_m }{|\omega_m|} \, .
\label{A1}
\end{equation}
The continuous-frequency version of (\ref{A1}) is
\begin{equation}
\lambda  \Delta (\omega ) = - g \int_{\alpha T <|x|<E_c} \, \frac{dx \Delta (x) }{2|x|} \,
\frac{(\omega-x)^2 + \Omega^2_1}{(\omega-x)^2 + \Omega^2} \, .
\label{A11}
\end{equation}

It is instructive to consider separately the case $\Omega_1 =0$, like in Bardeen-Pines model~\cite{BardeenPines},  and
the case when $\Omega_1$ is finite.

\subsubsection{The case $\Omega_1=0$}

We first solve for $T_c$, for which $\lambda_{\rm max} (T_c) =1$,
and then obtain $\lambda_{\rm max} (T)$.
To find $T_c$, we compare the gap equation at $\omega =0$ and at a finite $\omega$.
 Setting $\lambda =1$ in (\ref{A11}) yields
\begin{equation}
\Delta (\omega ) = - g \int_{\alpha T <|x|<E_c} \, \frac{dx \Delta (x) }{2|x|} \,
\frac{(\omega-x)^2}{(\omega-x)^2 + \Omega^2}  \, .
\label{A11_3}
\end{equation}
At zero frequency we have
\begin{equation}
\Delta (0) = - g \int_{\alpha T <x<E_c} \, \frac{dx \Delta (x) }{x} \,
\frac{x^2}{x^2 + \Omega^2} \, .
\label{A11_2}
\end{equation}
 Because the interaction vanishes at $x=0$, the lower limit of integration over $x$ can be safely set to zero.

 At a finite $\omega$, we single out logarithmically divergent term from $\int dx/x$ and write
 \begin{widetext}
 \bea
 \Delta (\omega) &=& \Delta (0) \left[1 - g (L + \ell) \frac{\omega^2}{\omega^2 + \Omega^2}\right] \nonumber \\
 && - g \int_{\alpha T <x<E_c} \frac{d x}{2x} \left\{ \Delta (x) \left[\frac{(\omega-x)^2}{(\omega-x)^2 + \Omega^2} + \frac{(\omega+x)^2}{(\omega+x)^2 + \Omega^2} - 2 \frac{x^2}{x^2 + \Omega^2}\right] -2 \Delta (0) \frac{\omega^2}{\omega^2 + \Omega^2}\right\} \,.
 \label{ch_3}
 \eea
 \end{widetext}
 One can easily make sure that the last term is free from infra-red singularity, hence the lower limit of the integration over $x$ can be set to zero.  This last term scales as $\omega^2$ at small $\omega \ll \Omega$ and reduces to $\Delta (0) g \ell$ at $\omega \gg  \Omega$.  When $\Omega \ll E_c$ and $\ell = \ln [E_c/\Omega]$ is large, it can be approximated, to logarithmic accuracy, by $g \Delta (0) \ell \omega^2/(\omega^2 + \Omega^2)$.
   It then cancels out the equivalent term in the first line in (\ref{ch_3}), and
   Eq.~(\ref{ch_3}) reduces to
  \begin{equation}
\Delta (\omega)  \approx \Delta (0) \left(1 - g L  \frac{\omega^2}{\omega^2 + \Omega^2}\right) \,.
\label{ch_4}
\end{equation}
Substituting this $\Delta (\omega)$
into the r.h.s. of (\ref{A11_2}), we obtain the equation on $T_c$:
\beq
1 = - g\ell + g^2 \ell L  \,.
\label{ch_5}
\eeq
Hence,
\beq
L = \ln {\frac{\Omega}{\alpha T}} = \frac{1+ g \ell}{g^2 \ell} \,.
\label{ch_6}
\eeq
One can verify that this is exactly the same expression as in Eq.~(\ref{ch_7}) at $f=1$, which in
Rietschel-Sham model corresponds to the vanishing of the repulsion at small frequencies.

Returning to (\ref{ch_4}), we see that the gap function
$\Delta (\omega)$
 changes sign at $\omega_c = \Omega/\sqrt{gL-1} = \Omega \sqrt{g \ell}$ and saturates at high frequencies to  $- \Delta(0)/(g \ell)$.

At $\ell = O(1)$, the analysis of $T_c$ becomes more involved as the prefactors for $g$ and $g^2$ terms in
(\ref{ch_6}) are determined by internal frequencies $x \sim \Omega$ rather than $E_F \gg x \gg \Omega$.
One way to proceed is to solve the gap equation (\ref{A11_3}) in direct perturbative expansion in $g$ (see Refs. \cite{Karakozov75,Dolgov05,Wang13,Marsiglio18}).
The computation is lengthy but straightforward. We skip the details and present the result:
\beq
L = \ln {\frac{\Omega}{\alpha T}} = \frac{1+ g \beta_1}{g^2\beta_2 - g^3 \beta_3} \,,
\label{ch_8}
\eeq
where
\bea
\beta_1 &=& \int_{0}^{E_c} \frac{ dx x}{x^2 + \Omega^2} = \ln {\frac{\sqrt{E^2_c + \Omega^2}}{\Omega}} \,, \nonumber \\
\beta_2 &=& \int_{0}^{E_c} \frac{ dx x^3}{(x^2 + \Omega^2)^2} = \ln {\frac{\sqrt{E^2_c + \Omega^2}}{\Omega}} - \frac{1}{2} \frac{E^2_c}{E^2_c + \Omega^2} \,, \nonumber \\
\beta_3 &=& \int_{0}^{E_c} \frac{ dx x^3 \Omega^2}{x^2 + \Omega^2} \Phi (x) \,, \nonumber \\
\Phi (x) &=&\int_{0}^{E_c} \frac{ dy y}{(y^2 + \Omega^2)^2}
\frac{ x^2 + \Omega^2 -3y^2}{(x^2+y^2 + \Omega^2)^2 - 4x^2y^2}   \,.
\label{ch_9}
\eea
At large $E_c/\Omega$, we have $\beta_1 \approx \beta_2 \approx 2\beta_3 \approx \ell$.
Keeping terms of order $g \ell$ but neglecting terms of order $g$, we reproduce Eq.~(\ref{ch_6}).  At $E_c \geq \Omega$, all three constants $\beta_i = O(1)$, and Eqs.~(\ref{ch_8}) and (\ref{ch_9})
yield the result for $T_c$ with corrections of order $g$.  Note that the leading term in the r.h.s. of (\ref{ch_8}) is $1/(g^2 \beta_2)$, hence the leading exponential dependence of $T_c$ is $T_c \propto E_c e^{-1/(g^2 \beta_2)}$.
We will see that the presence of $g^2$ in the exponent is specific to the case $\Omega_1 =0$.
For finite $\Omega_1$, the exponential factor contains $1/g$ rather than $1/g^2$.

\begin{figure}
\includegraphics[width=0.80\columnwidth]{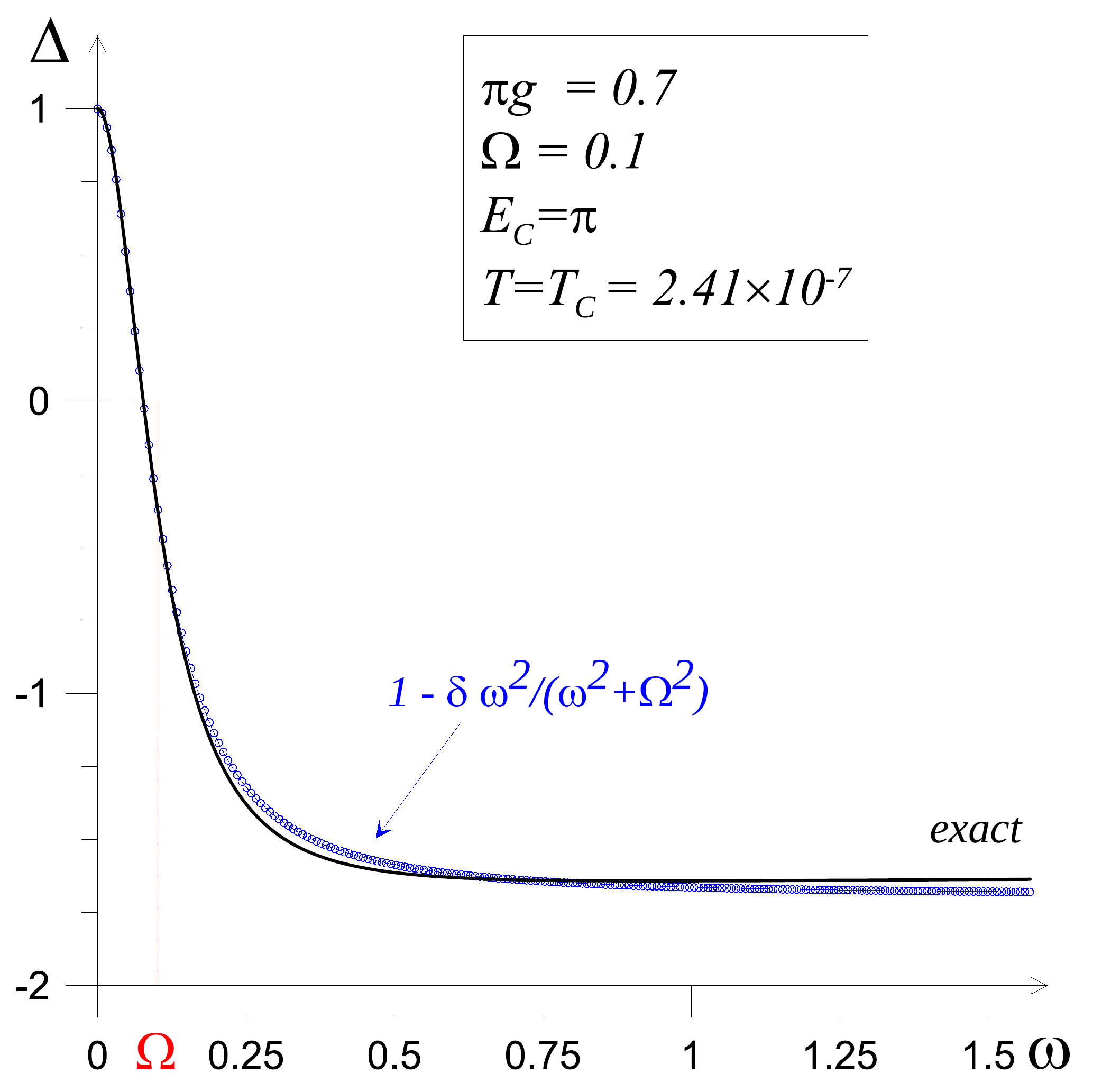}
\caption{(color online). Analytical (blue circles), Eq.~(\ref{ch_10}), {\it vs} numerical (black line) solutions
of Eq.~(\ref{A11}) for $T=T_c$.
\label{fig2}
}
\end{figure}

Eq.~(\ref{ch_8}) gives correct $T_c$ in the asymptotic limit $g \ll 1$, but at  realistic $g \leq 1$ higher order terms in $g$ may become substantial. To estimate $T_c$ at $\Omega \leq E_c$ and $g \leq 1$ we take as an input the
 numerical solution of the actual gap equation (\ref{A1}). It shows (see Fig.~\ref{fig2}) that the functional form of $\Delta (\omega)$ is consistent with  (\ref{ch_4}) to reasonably good accuracy,  but the prefactor for
 $\omega^2/(\omega^2 + \Omega^2)$ is different from $gL$.  We then search for the approximate solution
using the functional form
 \beq
 \Delta (\omega) = \Delta (0) \left(1 - \delta \frac{\omega^2}{\omega^2 + \Omega^2}\right) \;.
\label{ch_10}
\eeq
The parameter
 $\delta$ is determined self-consistently, by substituting (\ref{ch_10}) into the r.h.s. of (\ref{A11_3}),
expanding it in $\omega$, and matching the $\omega^2$ terms.  This yields
\begin{equation}
\delta = g\Omega^4 \int_{\alpha T}^{E_c} \frac{dx}{x}
 \, \frac{\Omega^2 - 3x^2}{(x^2 + \Omega)^3} \, \left( 1 - \delta \frac{x^2} {x^2 + \Omega^2 } \right)   \,.
\label{A7}
\end{equation}
Substituting further (\ref{ch_10}) into (\ref{A11_2}) we obtain
\begin{equation}
1 = - g \int_{0}^{E_0} dx \, \frac{ x}{x^2 + \Omega^2}
\left( 1 - \delta \frac{ x^2} {x^2 + \Omega^2} \right) \,.
\label{A5}
\end{equation}
Equations (\ref{A7}) and (\ref{A5})
determine $T_c$ and $\delta$ self-consistently.
The result is
\begin{equation}
\delta = \frac{1 + g\beta_1 }{ g \beta_2} \, .
\label{A6}
\end{equation}
and
\begin{equation}
L = \ln \left[\frac{\Omega}{\alpha T_c}\right]  =
c_1+\frac{(1 + g\beta_1) (1-c_2g)}{g^2\beta_2}
\,
\label{A8}
\end{equation}
where
\begin{equation}
c_1=\gamma_1+4\gamma_2+\ell-\beta_1 \,,
\label{A9}
\end{equation}
and
\begin{equation}
c_2=3\gamma_2-4\gamma_3 \,,
\label{A9b}
\end{equation}
are numerical coefficients of order unity (approaching, respectively, $3/2$ and $1/12$ in the $E_c/\Omega \gg 1$ limit),
and
\begin{equation}
\gamma_k=  \int_0^{E_c} \frac{ dx x \Omega^{2k}}{(x^2 + \Omega^2)^{1+k} } =
\frac{1}{2k} \left[ 1-\frac{\Omega^{2k}}{(E_c^2 + \Omega^2)^{k} } \right]  \,.
\label{A9c}
\end{equation}
For small $g$ and large $\ell = \ln [E_c/\Omega ]$, Eqs.~(\ref{A8}) and (\ref{ch_8}) agree to leading order,
but, predictably,  differ in corrections of order $g$.

To compare the two formulas, we estimate $T_c$ for the same parameters that were used in Fig.~\ref{fig2}.
Eq.~(\ref{A8}) yields $T_c =1.8 \times 10^{-7}$, and Eq.~(\ref{ch_8}) yields $T_c =3.3 \times 10^{-7}$.
The two values are quite close and also close to the numerical result $T_c = 2.41 \times 10^{-7}$.
This situation holds for smaller values of $E_c/\Omega$.
For the same $g=0.228$ and $E_c =\pi $ but  $E_c / \Omega =e$, Eq.~(\ref{A8}) yields $T_c =2.8 \times 10^{-18}$, and Eq.~(\ref{ch_8}) yields $T_c =4.4 \times 10^{-18}$, providing accurate estimates of the transition temperature
$T_c \approx 2 \times 10^{-18}$ obtained numerically.

The perturbative and the self-consistent computational schemes  can be easily extended to obtain temperature
dependence of the largest eigenvalue $\lambda_{\rm max} (T)$.
All one needs to do is to substitute $g$ with $g/\lambda(T)$ in Eq.~(\ref{A11_3}) and
re-express the condition for $T_c$ at $\lambda_{\rm max} =1$ as the
equation on $\lambda_{\rm max} (T)$. Within the self-consistent scheme we obtain
(for $E_c/\Omega \gg 1$)
\begin{equation}
\lambda_{\rm max} (T) = \frac{ g }{2} \left[ \sqrt{ 4 L (\ell-\frac{1}{2}) + \ell^2 - \frac{37}{6} \ell + \frac{433}{144} }
-\ell+\frac{1}{12} \right]   \,.
\label{A10}
\end{equation}
The dependence on $L$ is manifestly non-linear.

\subsubsection{Finite $\Omega_1$}

The computations at a non-zero $\Omega_1$ in $V_{\rm eff}$ in Eq.~(\ref{ch_1}) proceed in a similar way.
We skip the details and present the results. At large $\ell$, keeping powers of $g \ell$ but neglecting powers of $g$, we obtain
 \beq
 L = \frac{1 + g \ell}{g} \frac{\Omega^2}{g \ell \left(\Omega^2 - \Omega^2_1\right)- \Omega^2_1} \,,
 \label{ch_11}
 \eeq
 One can easily check that this is equivalent to Eq.~(\ref{ch_7}) for Rietschel-Sham model,
  if we identify $f$ with $1 - \Omega^2_1/\Omega^2$.  We see that there is a threshold on finite $T_c$
   at $\Omega_1/\Omega = [g \ell/(1+ g \ell)]^{1/2}$. For larger $\Omega_1$, frequency variation of the interaction is not sufficient, and $T_c =0$. For large enough $g\ell$, $T_c$ is finite.

  For $\Omega \leq E_c$, we again obtain $T_c$ in direct perturbative expansion in $g$. The perturbative analysis is only valid at small $\Omega_1/\Omega$, otherwise small $g$ are below the threshold.  Expanding in $g$ and in $\Omega_1/\Omega$, we obtain
  \beq
L = \ln \frac{\Omega}{\alpha T} = \frac{1+ g \beta_1}{g} \frac{1}{g\beta_2 - \frac{\Omega^2_1}{\Omega^2}- g^2 \beta_3} \,,
\label{ch_12}
\eeq
where $\beta_{i}$ are the same as in (\ref{ch_9}) -- corrections to $\beta_i$ due to non-zero $\Omega_1$ account for
the terms, which are smaller than the ones that we kept in (\ref{ch_12}).

\section{The numerical analysis of the eigenvalue problem and the breakdown of the logarithmic flow}
\label{sec:numerics}

We now discuss in more detail the numerical solution for $T_c$. Like we said in the Introduction, the
hallmark  of the weak-coupling BCS theory is logarithmic in temperature flow
of the largest eigenvalue, $\lambda(T)$, of the linearized gap-function equation (\ref{GE1}), see, e.g.,  \cite{Scalapino1969,Gladstone1969}.
We present it graphically in Fig.~\ref{fig1}(b) using the notion of the four-point vertex function $\Gamma$
irreducible in the Copper channel. If $\Gamma$ is replaced with a negative constant $-g/\rho_F$, where $\rho_F$ is the density of states on the Fermi-surface per spin component (i.e., if the interaction is attractive), then the eigenvalue increases with decreasing $T$ as
\begin{equation}
\lambda(T) = g\ln [\Omega /\alpha T]+ {\rm const} \,.
\label{log}
\end{equation}
Even if $T_c$ is extremely small, one can predict its value from higher temperature data by using linear in
$\ln [\Omega/\alpha T]$ extrapolation. Such an extrapolation appears to be  an indispensable part of any fully {\it ab initio}
approach in view of the technical challenge of explicitly dealing with the energies/frequencies ranging from the bandwidth
down to $T_c$.

However, in the case of frequency dependent repulsive interaction, the behavior of the largest eigenvalue $\lambda_{\rm max} (T)$ is not purely logarithmical with $T$, see Eq.~(\ref{solA}) for the Rietschel-Sham  model and Eq.~(\ref{A10}) for the model with $V_{\rm eff} (\omega)$. In both cases, the logarithmic scaling of $\lambda(T)$ fails because it was based on the implicit assumption that the eigenvector  does not depend on $T$, while in our case, the $a/b$ ratio in the Rietschel-Sham  model and the position of sign change of $\Delta (\omega)$ in the model with  $V_{\rm eff} (\omega)$ varies with temperature.

So far, all our considerations were done for momentum-independent $\Gamma$. However, the singular
part of the $G^{(2)}$-function is fully symmetric with respect to its frequency
dependence and the dependence on the distance to the Fermi surface.
An immediate conclusion then is that systems with substantial
dependence of the $s-$wave component of  $\Gamma$ on the magnitude of the momentum
will feature similar properties, if
a repulsive interaction is weakened near the Fermi surface. If this is the case, then even
a static repulsive interaction may give rise to a finite $T_c$
by exactly the same mechanism as the one we discussed above.

\subsection{Implicit renormalization approach}

At this point one might get an impression that the idea of numerically extracting $T_c$ from an eigenvalue/eigenvector problem in terms of genuine $\Gamma$ and  $G^{(2)}$ at $T\gg T_c$ is hopeless, and
the only meaningful way to proceed is the pseudopotential (explicit renormalization) approach (see Refs.~\cite{Scalapino1969,Gladstone1969,Morel1962}).
Nevertheless, it turns out that the eigenvalue/eigenvector problem in terms of
genuine $\Gamma$ can be reformulated in such a way that  the resulting eigenvalue
$\bar{\lambda}(T)$ does feature the desired simple logarithmic flow.
Moreover, $\bar{\lambda}(T)$ is essentially the eigenvalue of the renormalized problem
despite being obtained directly from the {\it genuine} $\Gamma$ without explicitly
constructing the pseudopotential counterpart for the latter.

\begin{figure}
\includegraphics[width=0.80\columnwidth]{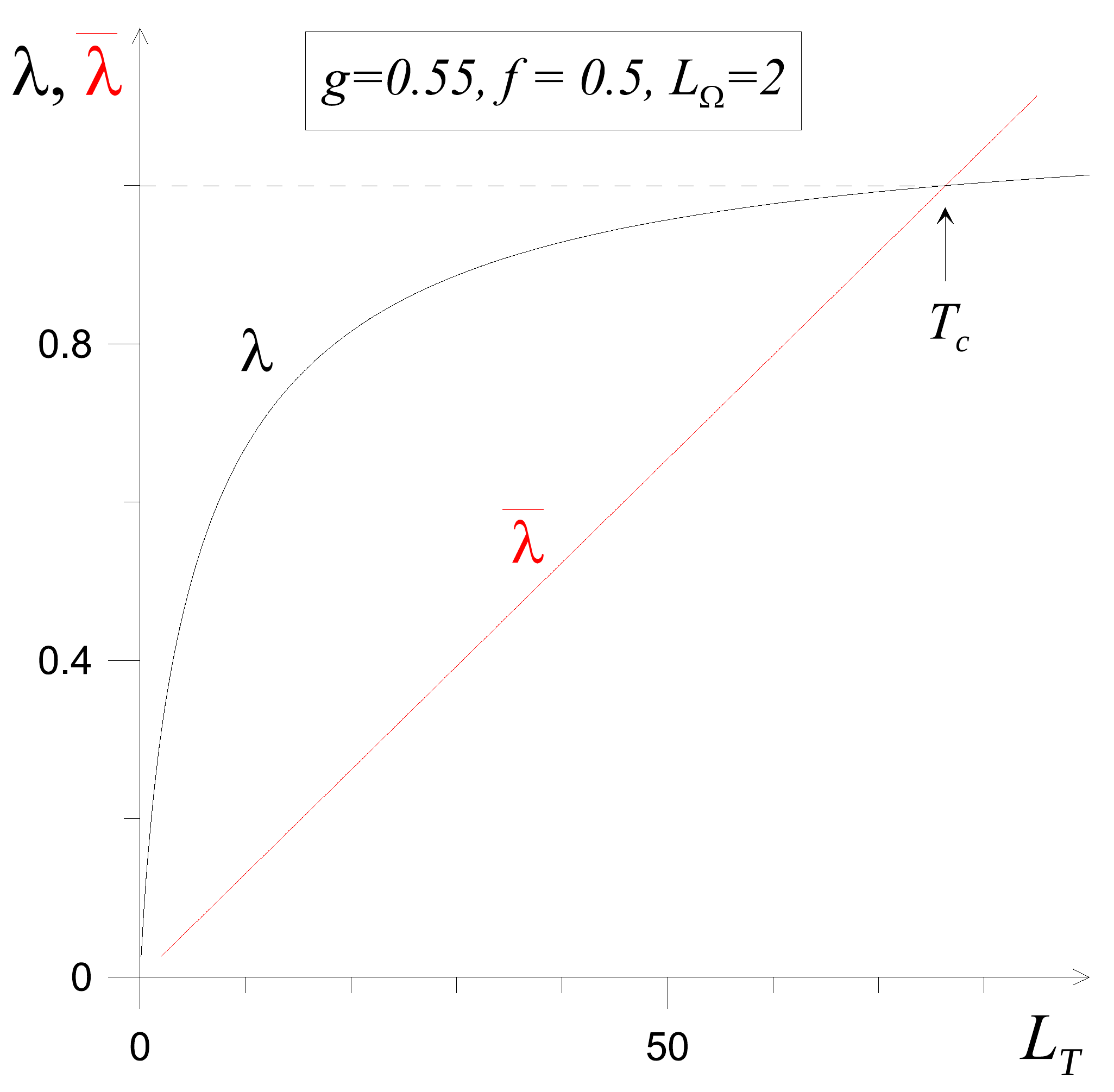}
\caption{(color online). The largest eigenvalue flow with logarithm of temperature for model (\ref{ModelA1})
when condition (\ref{Tolm}) is violated, Eq.~(\ref{solA}), (black line).
Flow of the modified eigenvalue problem is shown by the red line.
\label{fig3}
}
\end{figure}
\begin{figure}
\includegraphics[width=0.80\columnwidth]{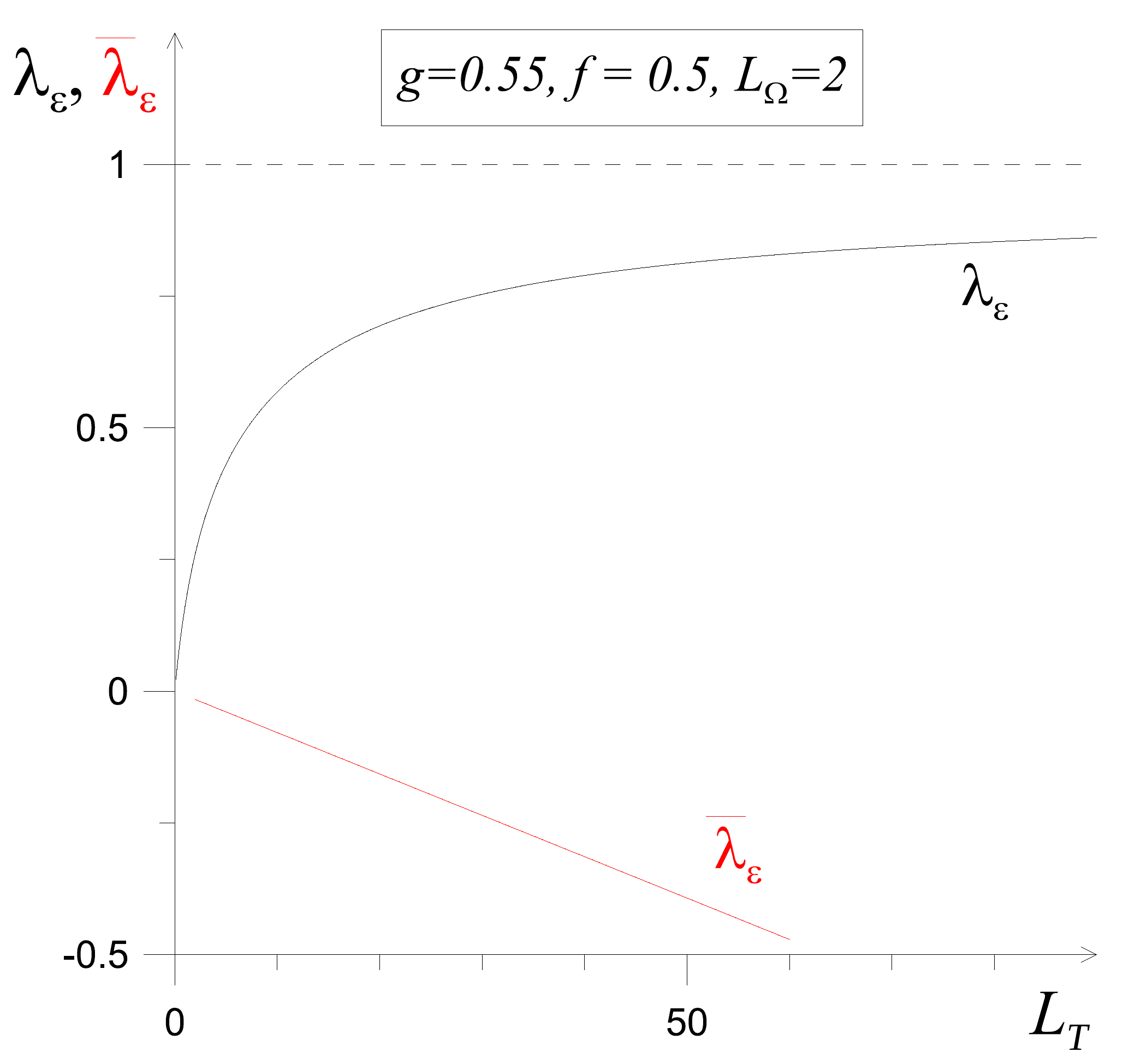}
\caption{(color online). Flow of $\lambda_{\epsilon}$ (black line) and
$\bar{\lambda}_{\epsilon}$ (red line) for $\epsilon =0.85$ when (\ref{Tolm}) is satisfied.
\label{fig4}
}
\end{figure}

\subsubsection{The Rietschel-Sham  model}

The Rietschel-Sham  model (\ref{ModelA1}) is perfectly suited for introducing our method, especially for tracing the intrinsic connection to---and the fundamental structural difference from---the pseudopotential approach.
First, let us change the parametrization of the gap function:
 \begin{equation}
\Delta = [ a\theta(|\omega_n|-\Omega) + s\theta(\Omega -|\omega_n|)]\theta(E_c-|\omega_n|) .
\label{repar}
\end{equation}
This way we separate $\Delta$ into two distinctively different parts: the high-frequency part, $|\omega_n| > \Omega$, (fully described by the parameter $a$) and the
the low-frequency part, $|\omega_n| < \Omega$, (fully described by the parameter $s$). With the new parameterization---but precisely the same eigenvector and eigenvalue, the problem (\ref{ModelA05}) is reformulated as
\begin{eqnarray}
\left\{ {\begin{array}{*{20}{c}}
 \lambda s= - g \ell a -g(1-f)L s ,\\
 \!\!\!\!\!  \!\!\!\!\!  \!\!\!\!\!  \!\!\!  \lambda a= -g \ell a -gLs  .
\end{array}} \right.
\label{sys_repar}
\end{eqnarray}
Make the formal replacement $\lambda \to 1$ in the second equation:
\begin{eqnarray}
\left\{ {\begin{array}{*{20}{c}}
\bar{\lambda} s= - g \ell a -g(1-f)L s ,\\
 \!\!\!\!\!  \!\!\!\!\!  \!\!\!\!\!   a= -g \ell a -gLs  .
\end{array}} \right.
\label{sys_new}
\end{eqnarray}
Now make a straightforward observation that the requirement $\bar{\lambda}(T_c)=1$ leads to precisely the same $T_c$---with precisely the same eigenvector $[a(T_c), s(T_c)]$---as the original system (\ref{sys_repar}). The utility of replacing (\ref{sys_repar}) with (\ref{sys_new}) becomes immediately clear by eliminating the high-frequency part $a$ between the two equations:
\begin{equation}
\bar{\lambda} s = g\left[ f-\frac{1}{1+g \ell} \right] L s  \;.
\label{pseudopot}
\end{equation}
The flow of $\bar{\lambda }$ is obviously linear in $L$. While the value of $T_c$ is, of course, independent
of the method, see Fig.~\ref{fig3}, the crucial difference is in the simplicity of extrapolating data
towards exponentially low temperature---in more complex models one may not be able to solve for eigenvalues
below $T \gg T_c$. For example, if we scale the coupling constant $g \to \epsilon g$ to smaller values to ensure
that the condition (\ref{Tolm}) is satisfied, the flow of $\lambda_\epsilon = \epsilon \lambda $ remains
qualitatively the same at $T \gg T_c$, see Fig.~\ref{fig4}. Reliable prediction of $T_c$ under these
conditions would be nearly impossible; moreover, one would be left wondering whether the model features
an effective attraction at low energies and ultimately goes SC.
In contrast, under the same conditions, negative $\bar{\lambda }_\epsilon$ would immediately signal
that the model is not SC in the corresponding channel.

By the very fact of eliminating the high-frequency part we understand that Eq.~(\ref{pseudopot}) corresponds to the peseudopotential theory with the
temperature flow of $\bar{\lambda}(T)$ controlled by the effective coupling constant $g_{\rm eff}$:
\begin{equation}
\bar{\lambda} =g_{\rm eff} L, \qquad
g_{\rm eff} =  gf-\frac{g}{1+g \ell}  .
\label{Tolm2}
\end{equation}

We indeed see that the structure of $g_{\rm eff}$ reproduces the Tolmachov-McMillan  logarithm \cite{Tolmachov2,Morel1962,Rietschel1983};
i.e. it demonstrates that a repulsive interaction is renormalized to a smaller value
$g/(1+g \ell)$ at low frequencies $(\omega_n, \omega_m) \lesssim \Omega$,
and SC instability is possible if this renormalized value is smaller than the bare low-frequency attractive term $fg$.
Diagrammatically, this result follows from the ladder summation for the effective low-frequency
$\Gamma$-function shown in Fig.~\ref{fig1}(c): $\Gamma_{\rm eff} = (g-gf) - g^2 \ell + g^3 \ell^2 - \dots$

While for the utterly simple Rietschel--Sham model (\ref{ModelA1}) the distinction between
the formulation (\ref{sys_new}) and the explicit pseudopotential formulation (\ref{pseudopot}) is merely
nominal, the difference becomes profound for any realistic case. Here the equivalents of the numbers $a$ and $s$ are the high- and low-frequency parts of (momentum- and frequency-dependent)  $\Delta$.  Correspondingly, the system (\ref{sys_new}) becomes the system of coupled integral equations (with the kernels given by genuine $\Gamma$).  At not too low temperature, the resulting eigenvector/eigenvalue problem remains solvable by techniques of DiagMC---even without explicitly evaluating $\Gamma$ (see Sec.~\ref{sec:newpseudo}). In contrast, the elimination of the high-frequency part required for going from (\ref{sys_new}) to (\ref{pseudopot}) would face the challenge of numerically constructing the vertex function $\Gamma_{\rm eff}$ out of the multi-variable and multi-scale vertex function $\Gamma$ (putting aside the non-trivial problem of
obtaining $\Gamma$ from first principles in a correlated  system).
\begin{figure}
\includegraphics[width=0.80\columnwidth]{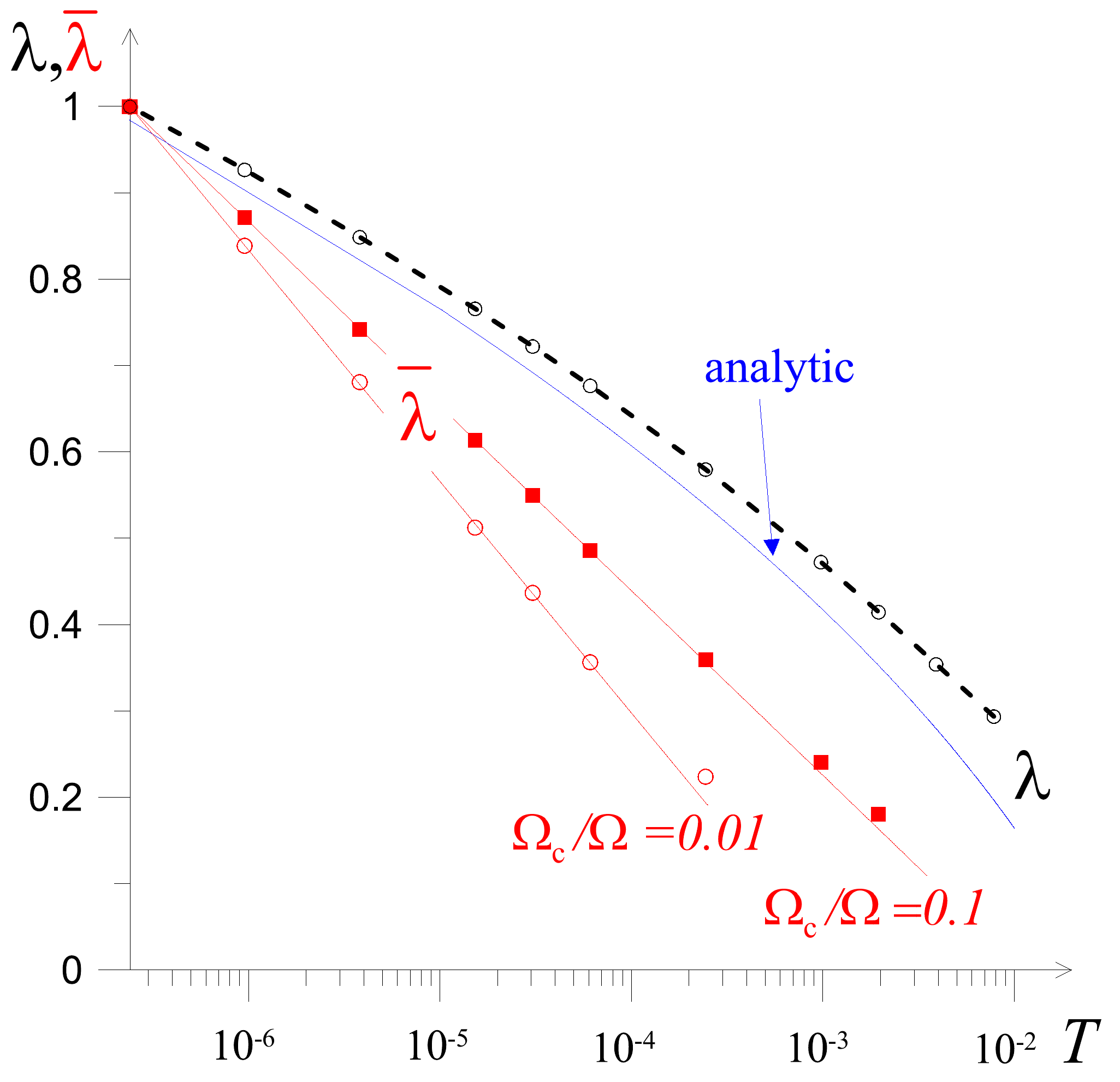}
\caption{(color online). Analytical solution (blue curve), Eq.~(\ref{A10}), of the eigenvalue problem (\ref{A11})
and the corresponding numerical solution (black circles) for the same model parameters as in Fig.~\ref{fig2}.
Filled squares and open circles show the numerical solutions of the same model
within the implicit renormalization approach with the energy scale separation set,
respectively, at $\Omega_c = 0.1 \Omega$ and $\Omega_c = 0.01 \Omega$.
\label{fig5}
}
\end{figure}

So far, we discussed the protocol of extracting $T_c$, but not the critical gap function. It turns out, however, that
the desired solution is immediately related to the eigenvector/eigenvalue problem of the implicit-renormalization scheme,
by simply multiplying the low-frequency part of the eigenvector,
obtained at a given temperature $T$, by the factor
$\bar{\lambda}(T)$. This relationship is readily traced with the model (\ref{ModelA1}) when the low-frequency part
of the solution is described by
 a single parameter $s$. We introduce the re-weighted quantity
\begin{equation}
s_*=\bar{\lambda} s
\label{s_bar}
\end{equation}
and
 re-express (\ref{sys_new}) in terms of $s_*$, thus arriving at the
eigenvalue/eigenvector problem for $(s_*,a)$:
\begin{eqnarray}
\left\{ {\begin{array}{*{20}{l}}
s_* = - g \ell a -(g/g_{\rm eff}) (1-f)  s_* ,\\
a= -g \ell a - (g/g_{\rm eff}) s_*  ,
\end{array}} \right.
\label{sys_new_bar}
\end{eqnarray}
where,
 by construction,
  $g_{\rm eff}=\bar{\lambda}(T)/L$. The  problem (\ref{sys_new_bar})
is free of any temperature dependence, meaning that the vector $(s_*,a)$ is temperature-independent.
 We now recall that at $T=T_c$, ${\bar \lambda} =1$, hence
 $s_*= s$, see Eq.~(\ref{s_bar}). This  implies
 that $(s_*,a)$ coincides with the critical gap function.

\subsubsection{The model with $V_{\rm eff} (\omega)$}

The same computational scheme can be use to obtain $T_c$ in the model with  $V_{\rm eff} (\omega)$, Eq.~(\ref{ch_1}).
The formulation of the implicit renormalization approach is as described above, but its practical implementation
is different for the final step.
Here we describe it using general vector-matrix notations (closely following Ref.~\cite{Rietschel1983}),
when the original $T_c$ problem can be written in a compact form as
\begin{equation}
\lambda \Delta_n = -\sum_m A_{n,m} \Delta_m  \;\;\; \to \;\;\;
\lambda \vec{\Delta } = -\hat{A} \vec{ \Delta }  \,.
\label{P1}
\end{equation}
First, the gap function is decomposed into two complementary parts (low- and high-frequency projections),
$\vec{\Delta} \equiv \vec{\Delta }^{(1)} + \vec{\Delta }^{(2)}$,
such that $\Delta^{(1)}_n =0$ for $|\omega_n|>\Omega_c$ and
$\Delta^{(2)}_n =0$ for $|\omega_n|<\Omega_c$. Correspondingly, the
$\hat{A}$-matrix is decomposed into four complementary matrixes,
$\hat{A}=\hat{A}^{(11)}+\hat{A}^{(22)}+\hat{A}^{(21)}+\hat{A}^{(12)}$,
such that $\hat{A}^{(11)}$ and $\hat{A}^{(22)}$
have zero matrix elements between low- and high-frequency subspaces, while
the only non-zero matrix elements of $\hat{A}^{(21)}$ and $\hat{A}^{(12)}$
are those connecting low-to-high and high-to-low frequency subspaces, respectively.
In analogy with (\ref{sys_new}), we then consider the eigenvector-eigenvalue problem
\begin{eqnarray}
\left\{ {\begin{array}{*{20}{c}}
\! \! \bar{\lambda} \vec{\Delta}^{(1)} = - \hat{A}^{(11)} \vec{\Delta}^{(1)} - \hat{A}^{(12)} \vec{\Delta}^{(2)} \,,\\
~\vec{\Delta}^{(2)} = - \hat{A}^{(22)} \vec{\Delta}^{(2)} - \hat{A}^{(21)} \vec{\Delta}^{(1)} \,.
\end{array}} \right.
\label{PP}
\end{eqnarray}
Along the same lines,  the relationship between this problem and the explicit renormalization approach is readily established by formally
substituting $ \vec{\Delta}^{(2)} = - [\hat{I}+\hat{A}^{(22)}]^{-1} \hat{A}^{(21)} \vec{\Delta}^{(1)}$ (the equality implied by the second equation) into the first equation.
This yields, in analogy with Eq.~(\ref{Tolm2}),
\begin{equation}
\bar{\lambda} \vec{\Delta}^{(1)} =  -\hat{B} \vec{\Delta}^{(1)} \,,
\label{P}
\end{equation}
where $\hat{B} = \hat{A}^{(11)} - \hat{A}^{(12)} [\hat{I}+\hat{A}^{(22)}]^{-1} \hat{A}^{(21)} $ is the renormalized
kernel in the Cooper channel. Its diagrammatic expansion in terms of the bare $\Gamma$ has the same structure as that of the series shown in Fig.~\ref{fig1}(c).

When the problem (\ref{PP}) is solved---the procedure is described in the next section---for model (\ref{A11})
with  $\Omega_c \lesssim \Omega$, the result is a nearly perfect
linear dependence of $\bar{\lambda }$ on $\ln \Omega_c/T$ at low temperature, see Fig.~\ref{fig5}.

\begin{figure}
\includegraphics[width=0.80\columnwidth]{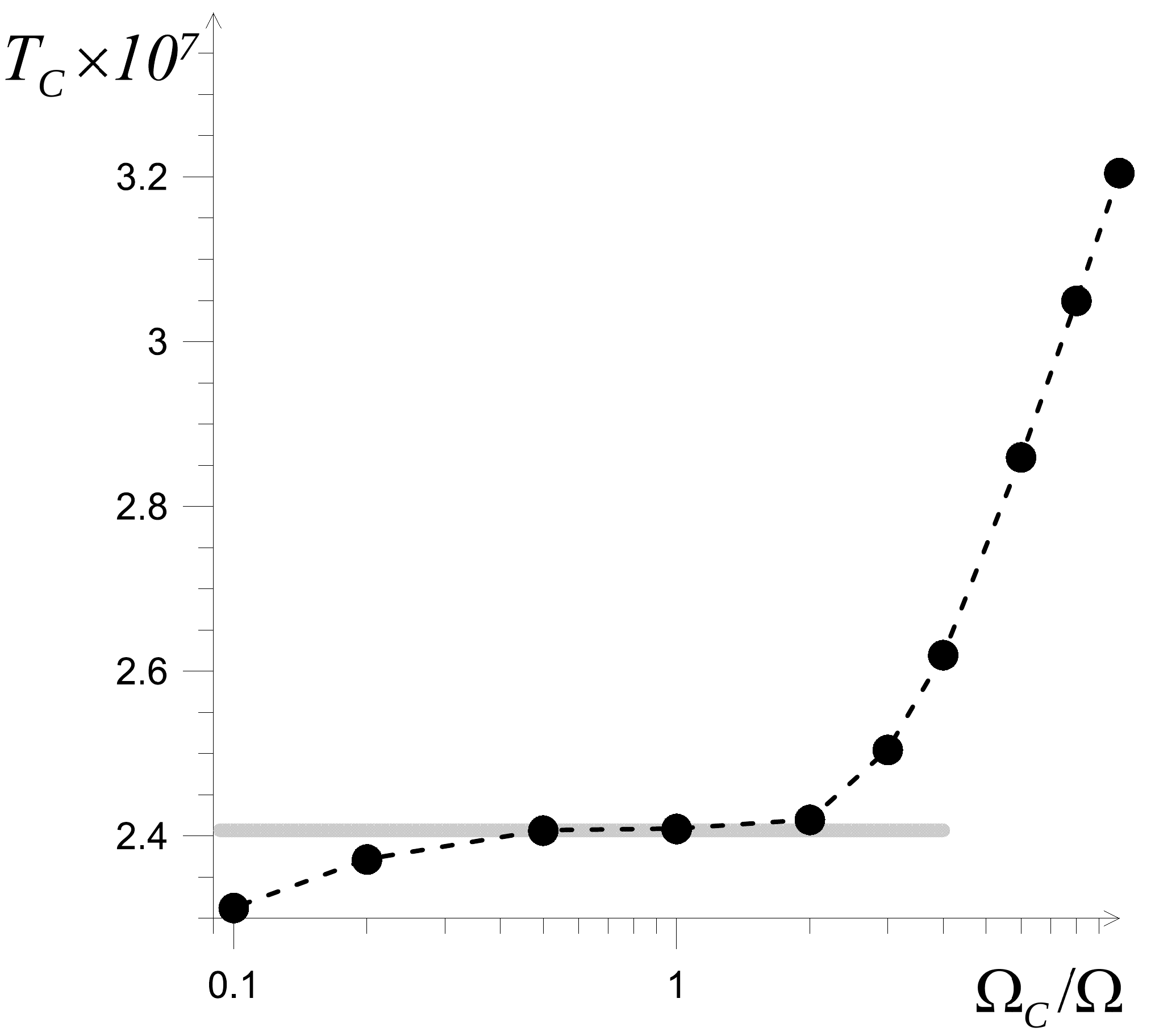}
\caption{Prediction of $T_c$ (for the same model parameters as in Fig.~\ref{fig2})
by linear in $\ln 1/T$ extrapolation of $\bar{\lambda }$
data from the temperature interval $10^{-5} < T < 10^{-4}$ using various high-frequency cutoffs $\Omega_c$.
The flow saturates to the correct value (bold grey line) only for $\Omega_c \lesssim \Omega$.
\label{fig6}
}
\end{figure}

In Fig.~\ref{fig5}, we also compare the numerical solutions for ${\bar \lambda}$ for the case $\Omega_1 =0$ to the analytical
solution, Eq.~(\ref{A10}), and the numerical solution of the
 the original eigenvalue problem, Eq.~(\ref{A11}), using
the same model parameters as in Fig.~\ref{fig2}.
We see that ${\bar \lambda} (T)$ is linear in $\ln T$ all the way down to $T_c$, and this allows one to determine $T_c$ in a controllable way by extrapolating ${\bar \lambda} (T)$ from higher temperatures.
 We illustrate this in Fig.~\ref{fig6}, where we show the results of the extrapolation using various high-frequency cutoffs $\Omega_c$.  We see that the extrapolation of  ${\bar \lambda} (T)$
yields the correct $T_c$  if we set $\Omega_c < \Omega$. On the other hand, we  clearly see from the Figure
that the non-linear dependence of the original $\lambda_{\rm max} (T)$ on $\ln T$
 results in a large overestimate of $T_c$, if the extrapolation is
done by using the original $\lambda_{\rm max} (T)$, or by setting the cutoff at $\Omega_c \gg \Omega$.
It is also clear that if we were to choose  $\Omega_c$ much smaller than $\Omega$, it would be
close to the extrapolation interval, and  this would result in increased systematic error.

Finally, we want to make sure that the vector $[\vec{\Delta}_*^{(1)} ,\, \vec{\Delta}^{(2)} ]$, wehre [cf.~(\ref{s_bar})]
\begin{equation}
\vec{\Delta}_*^{(1)} = \bar{\lambda} \vec{\Delta}^{(1)}\, ,
\label{Delta_1_star}
\end{equation}
yields the critical gap function. Rewriting the problem (\ref{PP}) in terms of $[\vec{\Delta}_*^{(1)} ,\, \vec{\Delta}^{(2)} ]$, we get
\begin{eqnarray}
\left\{ {\begin{array}{*{20}{c}}
 \vec{\Delta}_*^{(1)} = -[ \hat{A}^{(11)}/\bar{\lambda}] \, \vec{\Delta}_*^{(1)} - \hat{A}^{(12)} \vec{\Delta}^{(2)} \,,\\
\vec{\Delta}^{(2)} = - \hat{A}^{(22)} \vec{\Delta}^{(2)} - [\hat{A}^{(21)} /\bar{\lambda}] \, \vec{\Delta}_*^{(1)} \,.
\end{array}} \right.
\label{PP_bar}
\end{eqnarray}
The structure of this problem is similar to that of (\ref{sys_new_bar}), with an important distinction.
While in Eq.~(\ref{sys_new_bar}) the temperature dependence is
absent, Eq.~(\ref{PP_bar})
becomes {\it effectively} temperature independent only within the leading logarithmic accuracy,
when the logarithmic factor brought by the integration with
 the kernels $\hat{A}^{(11)}$ and $\hat{A}^{(21)}$
 is compensated by matching behavior
of $\bar{\lambda}(T)$. This means that $[\vec{\Delta}_*^{(1)} ,\, \vec{\Delta}^{(2)} ]$ reproduces
the critical gap function only
 within the leading logarithmic
  approximation. To recover an accurate result for the
critical gap function one has to
extrapolate $[\vec{\Delta}_*^{(1)} (T) ,\, \vec{\Delta}^{(2)} (T) ]$
to $T = T_c$.

\section{Implicit renormalization approach: techniques of implementation}
\label{sec:newpseudo}

\subsection{Iteration scheme}

Fully {\it ab initio} calculation of $T_c$ in metals
using Eq.~(\ref{PP}) faces two technical problems. To begin with, the vertex function has two
four-dimensional (momentum-frequency) indexes and its full tabulation, including high-energy scales
$\gg E_F$, is challenging; without simplifying assumptions the effort is about the square of what is
required for tabulating the Green's function $G_{{\mathbf p}, m}$. [Here we work with the extended
momentum space; otherwise ${\mathbf p}$ has to be understood as a composite label based on the
momentum in the first Brillouin zone and band index.] Suppose that this problem is taken care of.

Next, we need to solve the equation for $\vec{\Delta}^{(2)}$ [the second equation of the system (\ref{PP})] on a fine four-dimensional grid covering
all relevant energy scales from $\gg E_F$ to $\ll \omega_D$. At this point, inversion of the
$\hat{I}+\hat{A}^{(22)}$ matrix will pose a serious, if not unsolvable, problem because of the
huge matrix size (which can be reduced by employing coarse-graining description only
at the expense of accuracy and additional technical complexity). An attempt to solve
the equation by standard iterations,
\begin{equation}
\vec{\Delta}^{(2)}(i+1) = - \hat{A}^{(22)} \vec{\Delta}^{(2)}(i) - \hat{A}^{(21)} \vec{\Delta}^{(1)} \,,
\label{P4}
\end{equation}
will certainly fail because for Coulomb systems, the largest positive eigenvalues of $\hat{A}^{(22)}$
are very large as noted by Rietschel and Sham \cite{Rietschel1983}. The negative eigenvalues of $\hat{A}^{(22)}$
are all smaller than unity by the very statement of the problem---otherwise the system will go SC at
temperature above $\Omega_c$.

It turns out that the desired solution can be always found by a simple modification of the
iteration scheme. The idea follows from known convergence properties of ``damped" iterations, namely,
$f = - a f - b$ with $ a > -1$ can be solved by substituting into the r.h.s. the average of all
previous iterations:
\begin{equation}
f(i+1) = - a \frac{1}{i}\sum_{k=1}^{i} f(k)  -  b \,,
\label{P5}
\end{equation}
(see, for example, Ref.~\cite{welcome2007}). One can easily verify an extremely fast convergence of this
scheme even for large values of $a$. It is thus guaranteed that
\begin{equation}
\vec{\Delta}^{(2)}(i+1) = - \hat{A}^{(22)} \frac{1}{i}\sum_{k=1}^{i}  \vec{\Delta}^{(2)}(k)
- \hat{A}^{(21)} \vec{\Delta}^{(1)} \,,
\label{P6}
\end{equation}
will quickly converge to the desired solution, which in its turn can be subsequently used to
find the largest eigenvalue of (\ref{PP}) by the standard power method:
\begin{eqnarray}
&&\vec{\Delta}^{(1)}(j) = -\hat{B} \vec{\Delta}^{(1)}(j-1)  \,,   \;\;
\bar{\lambda }(j) = || \vec{\Delta}^{(1)}(j) || \,,  \nonumber \\
&& \vec{\Delta}^{(1)}(j) \to  \vec{\Delta}^{(1)}(j)/\bar{\lambda }(j) \, .
\label{P6_1}
\end{eqnarray}
Here $\hat{B} \vec{\Delta}^{(1)}$ is a short-hand notation
for the r.h.s. of the first equation of the
 set (\ref{PP})
  once we use for $\vec{\Delta}^{(2)}$ the solution of the second equation in (\ref{PP}).

In practice, we have two (internal and external) cycles and work with the original matrix $\hat{A}$.
In the internal (damped iterations) cycle, $\hat{A} \vec{\Delta}$ multiplication involves the full vector $\vec{\Delta}$
but the result is used to compute and average only the high-frequency components; the low-frequency part remains intact.
After convergence, the same $\hat{A} \vec{\Delta}$ multiplication is used in the external cycle
to obtain the new vector, estimate the eigenvalue $\bar{\lambda }$, and normalize the
low-frequency components. The norm $|| \vec{\Delta}^{(1)} ||$ can be defined in a number of ways,
say, through the inner product.

To improve efficiency, all equations should be projected on different symmetry channels, and their
largest eigenvalues be determined independently. This is important for correctly predicting the outcome
of close competition between two or more channels, because the order of the largest eigenvalues
can potentially change with temperature.

\subsection{Diagrammatic Monte Carlo method}

Within the  DiagMC approach, where statistics is accumulated and averaged as a matter
of principle, damped iterations are realized naturally \cite{welcome2007}. The internal cycle is automatically
realized by running a self-consistent scheme when statistics for high-frequency components of $\Delta $
is accumulated by sampling the diagrammatic space for the r.h.s., $ \Gamma G^{(2)} \Delta$, where
$\Delta$-function is known from averaging previously collected statistics.

The DiagMC approach also circumvents the difficulty of tabulating the vertex function explicitly.
Formally, the $\Gamma$-function is represented by the series of Feynman diagrams and thus its
evaluation involves summation over diagram orders and topologies, as well as multi-dimensional
integrals and sums over internal momenta and frequencies. In DiagMC, all diagrammatic space
parameters, including external variables, are sampled stochastically without systematic bias.
In this sense, the r.h.s. of Eqs.~(\ref{PP}) are subject to the same DiagMC
simulation as electron self-energy or polarization function. At no point one has to worry about
handling an object more complex than the single-particle Green's function.
In fact, the gap function is more simple than $G$ because it lacks singular
momentum and frequency dependance, see Fig.~\ref{fig2}.

\section{Discussion and Conclusions}
\label{sec:conclusions}

In this communication we considered superconductivity in systems where the interaction is repulsive, but depends on  frequency, and is weaker at smaller frequencies than at larger ones.  This is a typical systems in a metal, where the interaction in the particle-particle channel is a combination of a stronger repulsion due to Coulomb interaction, and a weaker attraction due to electron-phonon interaction.

There are at least three characteristic energy scales in metals:
the Fermi energy $E_F$, the
plasmon frequency $\omega_p$, and the Debye frequency $\Omega_D$.
Screening of the Coulomb interaction develops
at frequencies below $\omega_p$, but for small momenta $q \ll k_F$ it is not complete down to
 $\omega \leq v_F q$, where $v_F$ is the Fermi velocity.
The electron-phonon interaction
becomes important at $\omega \lesssim \Omega_D$.
In the conventional pseudopotential approach, all effects of repulsive Coulomb interaction, including its momentum
and frequency dependence, are absorbed into a single semi-phenomenological
dimensionless coupling $\mu_*$
(see Ref. ~\cite{Giustino2017} for a recent review on this issue).  Superconductivity develops if a
 dimensionless $\lambda_{ph}$ due to electron-phonon interaction exceeds $\mu^*$.

In our analysis of superconductivity  somewhat different approach, inspired by recent studies of  $s$-wave superconductivity in  SrTiO$_3$ and Bi. Namely, we assumed, following earlier studies~\cite{Mahan,Ruhman2016,Ruhman2017,Maria2019}, that the effective interaction in the particle-particle channel can be viewed as the  sum of a screened Coulomb interaction and an interaction with a gapped boson, which is a hybridized mode between a longitudinal phonon and a plasmon.  We considered s-wave superconductivity and focused on the frequency dependence of the effective pairing interaction, and neglected its momentum dependence.

 We considered the two models: the Rietschel-Sham model with a step-like pairing interaction, and the model with a continuous pairing interaction $V_{\rm eff} (\omega)$ (Eq.~(\ref{ch_1})).  In both models the pairing interaction reduces to a larger constant $g$ at high energies and to a smaller constant $g(1-f)$ at
 the lowest energies.  We found that $T_c$ is finite already at arbitrary small $g$, if $f=1$, i.e., if the interaction vanishes at the smallest frequencies. However, if $f$ is smaller than one, $T_c$ is finite only if $g$ exceeds a certain threshold.  The threshold values and the  results for $T_c$ for the two models are similar, but not equivalent. For the model with a continuous interaction $V_{\rm eff} (\omega)$ we computed $T_c$ using various computational schemes,  and analyzed the interplay between the threshold value and the ratio $\Omega/E_c$, where $\Omega$ is the boson frequency and $E_c \leq E_F$ is the frequency, up to which one can expand the fermionic dispersion to linear order in $k-k_F$.  We assumed that $\Omega < E_c$ and showed that the threshold value is reduced when  $\Omega/E_c$ gets smaller.

We also discussed in all detail the protocol for numerical computation of $T_c$ for systems with frequency dependent repulsive interaction, particularly for the cases when $T_c$ is small, and one needs to extrapolate the results for the largest eigenvalue of the gap equation, $\lambda_{\rm max} (T)$,
from temperatures $T \gg T_c$ to $T = T_c$.
 We demonstrated that
 within the standard setup, used for the systems with frequency independent attractive interaction,
 such an extrapolation is not possible due to
non-linear dependence
 of $\lambda_{\rm max} (T)$ on the logarithm of temperature. This non-linear dependence emerges  because
frequency-dependent repulsive interaction undergoes
significant changes at intermediate energy/momentum scales.

To set the new protocol, we noticed that the numerical calculations necessary involve the energy cutoff at $\Omega_c$.  In most calculations, this cutoff is set at $\Omega \ll \Omega_c \ll E_F$.  This is made
for purely technical reasons---to have better momentum resolution near the Fermi surface.
However, with this choice of $\Omega_c$ one faces the problem of extrapolating non-linear in
$\ln (\omega_D/T)$ data towards low temperature, if $T_c$ happens to be much smaller than the lowest possible temperature in the calculation. Another, more general drawback, is that for $\Omega_c \ll E_F$, the effects of Coulomb interactions are no longer included at the fully {\it ab initio} level and the subsequent calculation contains an unknown systematic error, not to mention that momenta satisfying $q < \Omega_c / v_F $ are treated inadequately. Moreover,  in multi-band or strongly anisotropic
systems the effects of Coulomb interaction cannot be described by a single parameter.

We argued that
 accurate evaluation of $T_c$ from Fermi-liquid properties at $T \gg T_c$
in correlated systems, where the BCS regime is an emergent phenomenon involving multiple energy scales,
cannot be achieved unless the renormalization scale $\Omega_c$ is made smaller (much smaller)
than $\Omega$.
This choice, however, brings about two technical problems when it comes to the practical implementation
of the method following the protocol described in Ref.~\cite{Rietschel1983}.
One is the necessity to know the full vertex function in a broad frequency and momentum range; the amount of
information is about the square of that required for knowing the single-particle Green's function.
Even if $\Gamma^{{\mathbf k},n}_{{\mathbf p}, m}$ can be tabulated without approximations and
any loss of accuracy, solving for high-frequency components of the gap function by matrix inversion
would be impossible because of the huge matrix size. Finding the solution by standard iterations
will not work either because negative eigenvalues for the full problem are largest in modulus
and for Coulomb systems will exceed unity already at $\omega \ll T \ll E_F$ \cite{Rietschel1983}.

A protocol for extrapolating numerical data towards $T_c$ from higher temperatures---applicable to
first-principle description of real metals---has to adequately capture the physics of the emergent
weakly-interacting effective theory. We have formulated the so-called  implicit renormalization
approach and demonstrated that it provides a simple, efficient, and unbiased protocol
for solving the extrapolation problem. The scheme has a built-in tool of controlling the systematic
error of extrapolation (see Fig.~\ref{fig6})---the only systematics of the otherwise numerically exact method.
The implicit renormalization approach is perfectly compatible with the diagrammatic Monte Carlo techniques,
allowing one to solve the corresponding eigenvalue problem without invoking the matrix inversion or
even explicitly calculating the four-point vertex function $\Gamma$.
 The  implicit renormalization protocol also allows one to obtain the correct gap function immediately below $T_c$.

Throughout the paper, the separation of the gap function into the low-energy part $\vec{\Delta}^{(1)}$ and the higher-energy part $\vec{\Delta}^{(2)}$
was performed in the  frequency domain.
Our approach, however, can be readily extended to include this separation for
 both the frequency and momentum variables.
  For example, the condition
 on the ``low-energy" regime, where $\vec{\Delta} ({\bf p}, \omega) = \vec{\Delta}^{(1)}({\bf p}, \omega)$ can be as simple as
$\xi_p^2 + \omega^2 \leq \Omega_c^2$. Outside this range,
$\vec{\Delta}({\bf p}, \omega) = \vec{\Delta}^{(2)}({\bf p}, \omega)$.

\begin{acknowledgments}
We thank M. Gastiasoro, R. Fernandes,  D. Maslov, and A. Millis for fruitful discussions.
The work by AVC  was supported by the Office of Basic Energy Sciences U. S. Department of Energy under award DE-SC0014402.
\end{acknowledgments}

\bibliography{bibliography_cps}
\end{document}